\documentclass[a4paper,12pt]{article}
\usepackage[round]{natbib}
\usepackage[english]{babel}
\usepackage{setspace}
\usepackage[utf8x]{inputenc}
\usepackage{graphicx} % Allows including images
\usepackage[left=2cm,right=2cm,top=2cm,bottom=2cm]{geometry}
\usepackage[dvipsnames]{xcolor}
\usepackage{textcomp}
\usepackage[T1]{fontenc}
\usepackage{amssymb}
\usepackage{xcolor}
\usepackage{authblk}
\usepackage{ulem}
\usepackage{hyperref}
\usepackage{lineno}
\usepackage{appendix}
\usepackage{float}
\usepackage{upgreek}
\usepackage{multirow}
\title{
\large
\textbf{Eruption column modelling of explosive volcanism on Venus}
}

\author[1]{Maxence Lef{\`e}vre}
\author[2]{Matteo Cerminara}
\author[3]{Antonio Costa}
\affil[1]{LATMOS/IPSL, Sorbonne Universit\'e, UVSQ, Universit\'e Paris-Saclay, Centre National de la Recherche Scientifique, Paris, France}
\affil[2]{Istituto Nazionale di Geofisica e Vulcanologia, Pisa, Italy}
\affil[3]{Istituto Nazionale di Geofisica e Vulcanologia, Sede di Perugia, Perugia, Italy}
\date{Published in JGR: Planets, 130, e2025JE009320. \url{https://doi.org/10.1029/2025JE009320}}

\begin{document}

\maketitle
%\newpage
\begin{abstract}
%<250 words
Volcanism on Venus has never been directly observed, but several measurements indicate present-day activity. Volcanism could potentially play a role in climatic processes on Venus, especially in the sulfur cycle like on Earth. Observation of volcanic activity is the primary objective of future Venus spacecraft. However, there are many unknowns regarding its Venusian characteristics, like the condition at the vent, the volatile content and composition. Past modelling efforts have only studied explosive volcanic plume propagation over a limited range of flow parameters at the vent and in an idealised Venus atmospheric configuration. We propose to use the 1D FPLUME volcanic plume model in a realistic Venusian environment. In similar Venusian conditions, the height of the plume is consistent with past modelling. The present study shows that explosive volcanism would preferably reach 15~km of altitude. Under certain conditions, plumes are able to reach the VenSpec-H tropospheric altitude range of observations and even the 45~km cloud floor. For the first time, the impact of wind was quantified, and the super-rotating winds have a substantial impact by plume-bending of reducing the height of plumes. Contrary to the Earth, the atmospheric heat capacity depends greatly on temperature, and will disadvantage lower plumes and allow larger plumes to propagate at higher altitudes. The high latitude atmospheric environment, due to the thermal profile and weaker winds, is favorable to plumes reaching higher altitudes.
\end{abstract}

\section*{Plain Language Summary}
%<200 words
On Earth, volcanic outgassing has a great impact on the climate, playing a key role in the sulfur cycle, injecting ash that can possibly have a significant radiative impact. On Venus, the contribution of such injections to the climate history is not known. Recent space missions gave evidence of present-day activity, but no volcanic plume has ever been observed directly, and its characteristics are not known. This is the primary objective of future space missions. Past modelling efforts showed that under certain conditions, plumes could reach the 45~km cloud bottom. However, they have only studied explosive volcanic plume propagation over a limited range of flow parameters at the vent, and in an idealised Venus atmospheric configuration. In this study, we propose to use an internationally recognised eruption column model built for Earth studies and to adapt it to a realistic Venusian environment. The present study shows that explosive volcanism would preferably reach 15~km of altitude, and a sensitive study of plume and atmospheric parameters was conducted. Under certain conditions, plumes can reach the future EnVision spectrometers tropospheric altitude range of observations and even the 45~km cloud floor and above. The windshear and latitudes have the biggest impact.

%%%%%%%%%%%%%%%%%%%%%%%%%%%%%%%%%%%%%%%%%%%%%%%
%
%  BODY TEXT
%
%%%%%%%%%%%%%%%%%%%%%%%%%%%%%%%%%%%%%%%%%%%%%%%

%%% Suggested section heads:
% \section{Introduction}
%
% The main text should start with an introduction. Except for short
% manuscripts (such as comments and replies), the text should be divided
% into sections, each with its own heading.

% Headings should be sentence fragments and do not begin with a
% lowercase letter or number. Examples of good headings are:

% \section{Materials and Methods}
% Here is text on Materials and Methods.
%
% \subsection{A descriptive heading about methods}
% More about Methods.
%
% \section{Data} (Or section title might be a descriptive heading about data)
%
% \section{Results} (Or section title might be a descriptive heading about the
% results)
%
% \section{Conclusions}

\section{Introduction}

On Earth, volcanic eruption plays a key role in climate and sulphur chemistry, with episodic stratospheric injection. On Venus, 85,000 volcanic edifices have been identified \citep{hahnMorphologicalSpatialAnalysis2023}, ranging from less than 5 km to over 100 km in diameter. By scaling Earth’s volcanic rate to Venus' geomorphology, the number of active volcanoes would be between 40 and 120 per Earth year \citep{byrneEstimatesFrequencyVolcanic2022,vanzelstCommentEstimatesFrequency2022}. The question of volcanism is central to determine the evolution of the atmosphere and whether there ever was an ocean at the surface \citep{gillmannLongTermEvolutionAtmosphere2022}. The climatic impact of volcanic outgassing on Venus is not known, but the sulphuric acid cloud deck, between 47 and 70~km, could be maintained through a supply of sulphur dioxide gas \citep{bullockRecentEvolutionClimate2001}. Volcanic plume destabilising the cloud chemistry is the main hypothesis for the periodical variations of mesospheric SO$_2$ \citep{espositoSulfurDioxideEpisodic1984,marcqVariationsSulphurDioxide2013}. 

Magellan data have provided very little evidence of explosive volcanism \citep{headVenusVolcanismClassification1992}. Recent analysis of images from the Magellan mission showed geomorphological variations over a few months attributed to volcanism \citep{herrickSurfaceChangesObserved2023,sulcaneseEvidenceOngoingVolcanic2024}, and hot spots have been observed, suggesting possible present volcanic activity \citep{smrekarRecentHotspotVolcanism2010,shalyginActiveVolcanismVenus2015}. Nevertheless, the characteristics of volcanic plumes on Venus remain unknown \citep{wilsonPossibleEffectsVolcanic2024}, particularly the potential for explosive plumes \citep{headiiiVolcanicProcessesLandforms1986}, given the extreme planet's surface atmospheric density, which is approximately 65 times larger than on Earth. Only a few suspected pyroclastic features have been found on Venus from Magellan radar observations \citep{ganeshRadarBackscatterEmissivity2022}, typical of explosive flow.
\bigbreak
Due to scarce data, there was only a limited number of studies on Venus' explosive volcanic plumes. Numerical models were developed to study the buoyancy of volcanic plumes in such an environment  \citep{thornhillTheoreticalModelingEruption1993,kiefferNumericalModelsCalderaScale1995,robinsonLargescaleVolcanicActivity1995,glazeTransportExplosiveVolcanism1999,carazzoRiseTurbulentPlumes2008,glazeExplosiveVolcanicEruptions2011}. By testing a limited part of the vent flow parameters (mass flow rate, velocity and temperature), these studies concluded that volcanic plumes on Venus are unable to reach  70~km cloud-top altitude. The higher atmospheric surface pressure and temperature would tend to lower the ejection velocities compared to Earth and reduce the relative density difference between the volcanic mixture and the environment \citep{wilsonComparisonVolcanicEruption1983}, resulting in the formation of preferably pyroclastic fountains. \cite{aireyExplosiveVolcanicActivity2015} investigated the conditions that will promote explosive volcanic activity on Venus, and simulated conduit processes, concluding that the addition of CO$_2$ to an H$_2$O-driven eruption will increase the final pressure, velocity, and volume fraction of gas.
\bigbreak
The study of the exotic, hot, sulphur cycle of Venus will be exceptional in the next decade, with the observations of active volcanism the primary target. The ESA’s EnVision, and two NASA missions, DAVINCI \citep{garvinRevealingMysteriesVenus2022} and VERITAS, were all three selected to launch around 2031. A  spectrometer suite will be onboard EnVision, consisting of three completely independent channels, VenSpec-M \citep{helbertVenusEmissivityMapper2020} and VenSpec-H \citep{neefsVenSpecHSpectrometerESA2025} in the IR and VenSpec-U in the UV \citep{marcqInstrumentalRequirementsStudy2021}. These three channels will perform gas measurements from the surface to the cloud top. VenSpec-M will measure nightside H$_2$O column densities of the lowest 10~km. VenSpec-H will measure both on the nightside H$_2$O in the first 10~km and the H$_2$O, CO, OCS, SO$_2$ and HF gas ensemble between 30 and 45~km. It will also probe the H$_2$O, CO, OCS, SO$_2$ and HF gas ensemble between 65 and 80~km on the dayside. VenSpec-U will measure SO$_2$ and SO between 65 and 75~km on the dayside. The combined monitoring for the first time of gas over 70~km could be able to measure a gas increase due to a volcanic plume, and assess its composition and potentially also the composition of the magma.
\bigbreak
In this study, we aim to understand the vertical propagation of an explosive volcanic plume and its sensitivity in the hot and dense Venusian atmosphere to provide estimates of source conditions for future observations. To perform this study, we use a proven eruptive column model in the most realistic Venus environment so far. The impact of the wind shear and the dependency on temperature of the atmospheric specific heat are taken into account for the first time.
\bigbreak
Our paper is organised as follows. The model is described in Section~\ref{Model}. In Section~\ref{Comp}, the results are compared to the literature with a similar configuration. The propagation of a volcanic plume in a realistic Venusian environment is investigated in Section~\ref{Real}. The plume vertical propagation is discussed in Section~\ref{Vert}.
The sensitivity to several parameters is quantified in Section~\ref{sens}, and our results are discussed in Section~\ref{Disc}. Our conclusions are
summarised in Section~\ref{Conc}.

\section{Modeling}
\label{Model}
\subsection{Eruption column model}
We use the FPLUME model \citep{folchFPLUME10IntegralVolcanic2016,macedonioUncertaintiesVolcanicPlume2016} based on the solution of the equations for the conservation of mass, momentum, and energy in terms of cross-section averaged variables \citep{woodsFluidDynamicsThermodynamics1988,bursikEffectWindRise2001}. Particle fallout, particle re-entrainment, entrainment of ambient moisture, phase changes of water, and wind profile are taken into account by the model. Turbulent entrainment of ambient air plays an important part in the dynamics of jets and buoyant plumes. It is parameterised using two dimensionless coefficients that control the entrainment along the stream-wise (shear) and cross-flow (vortex) directions. The formulations of \cite{carazzoRouteSelfsimilarityTurbulent2006} and \cite{tateRiseDilutionBuoyant2002} are used respectively for the shear and vortex direction. The region above the Neutral Buoyancy Layer (NBL) is resolved using a semi-empirical approach, assuming pseudo-gas relationships with pressure equal to the atmospheric pressure at each level, and an adiabatic cooling temperature decrease. The FPLUME is an internationally recognised model, with performance for Earth's volcanic plume in line with the literature \citep{costaResultsEruptiveColumn2016}. This is the first use of the FPLUME in an extraterrestrial environment. 

\subsection{The Venus Environment}

The model was modified by setting the acceleration of gravity, the atmospheric molecular mass and heat capacity to Venus values, and the viscosity of the atmosphere set to CO$_2$ values \citep{whiteViscousFluidFlow2022}.

\begin{table}[!ht]
\center
\begin{tabular}{lcc}
%\hline
 & Venus & Earth \\
\hline
Acceleration gravity g (m~s$^{-2}$) & 8.87 & 9.81  \\
Surface pressure ($10^5$ Pa) & 92 & 1.013\\
Surface temperature (K) & 730 & 288\\
Molecular mass (g~mol$^{-1}$) & 43.4 & 28.9 \\
Surface density (kg~m$^{-3}$) & 65 & 1.23 \\
\hline
\end{tabular}
\smallbreak
\caption{Comparison of atmospheric characteristics between the Earth and Venus}
\label{T1}
\end{table}

\begin{figure}[!ht]
 \centering
   \includegraphics[width=8cm]{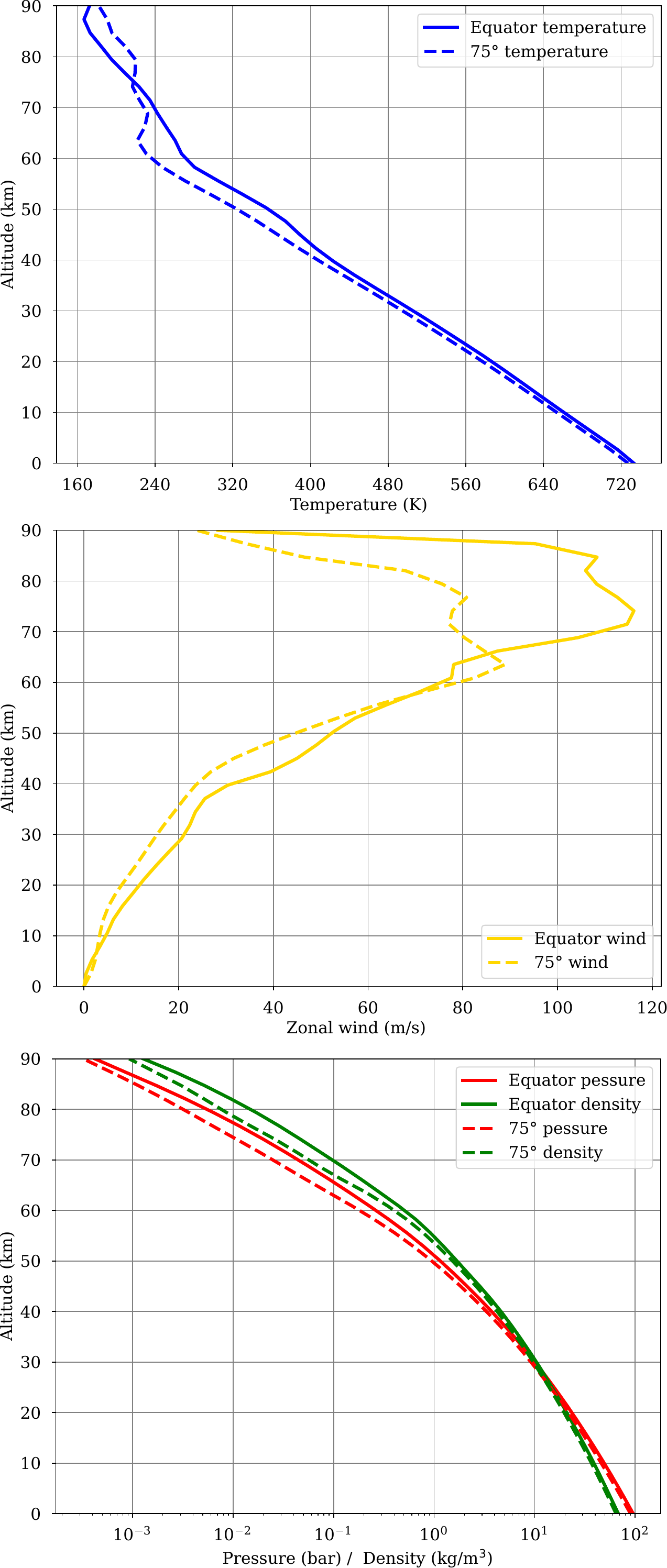}
    \caption{Vertical profiles of the Venus atmospheric temperature (top) in K, zonal wind (middle) in m~s$^{-1}$, and pressure (bottom) in bar, density (bottom) in kg/m$^3$ and for the Equator (solid lines) and 75$^{\circ}$ of latitude (dashed lines). The data comes from The Venus Climate Database.} 
  \label{21}
\end{figure}

As input, vertical profiles of the temperature, pressure, zonal winds and density extracted from The Venus Climate Database (Figure~\ref{21}) are prescribed to the model. The Venus' environment differs from the Earth by a hot and dense troposphere, and increasing winds up to 80~km. On Earth, the latent heat released as the entrained water vapour condenses may provide additional buoyancy \citep{woodsMoistConvectionInjection1993}. However, in Venus' troposphere, the amount of water is orders of magnitude below the saturation mixing ratio \citep{marcqCompositionChemistryNeutral2018}. The specific humidity (not shown here) is around 10$^{-2}$ g/kg at the surface and decreases to 10$^{-3}$ g/kg at 90~km. Therefore, no condensation processes are taken into account in this study.

The values of the mass flow rate (MFR hereinafter), exit temperature and exit velocity are set respectively between 10$^4$ and 10$^{11}$~kg~s$^{-1}$, 1000 and 1600~K and 100 and 340~m~s$^{-1}$. The majority of explosive eruptions, with a Volcanic explosivity index (VEI) below 7, correspond to MFR below 10$^{9}$~kg~s$^{-1}$ \citep{mastinMultidisciplinaryEffortAssign2009}, above this threshold, these are Ultra-Plinian eruptions occurring 1-2 times per thousand years on Earth \citep{newhallAnticipatingFutureVolcanic2018}. Above an MFR of 10$^{9}$~kg~s$^{-1}$, the eruptions are considered as explosive super-eruptions \citep{costaUnderstandingPlumeDynamics2018}. From the volume of the Scathach Fluctus pyroclastic flow deposit \citep{ghailPyroclasticFlowDeposit2015}, \cite{aireyExplosiveVolcanicActivity2015} estimated an MFR superior to 3$\cdot$10$^{8}$~kg~s$^{-1}$. \cite{wilsonComparisonVolcanicEruption1983} also estimated the MFR of a stable plume between 3~10$^{7}$ and 2~10$^{9}$~kg~s$^{-1}$. For clarity, only plumes with an MFR between 10$^7$ and 10$^{10}$~kg~s$^{-1}$ will be shown in the main text. \cite{fagentsExplosiveVolcanicEruptions1993} estimated the exit velocity from the motions of ballistic bombs for explosive eruptions to be between a few tens of m~s$^{-1}$ to 400~m~s$^{-1}$. Typical exit temperatures at the vent are between 900 and 1600~K \citep{lesherChapter5Thermodynamic2015}, the range of estimated Venusian values of \cite{aireyExplosiveVolcanicActivity2015} lies within it. Due to its high temperature and composition, the sound speed on Venus is reaching 410~m~s$^{-1}$ \citep{petculescuAtmosphericAcousticsTitan2007}, 20\% higher than on Earth, allowing higher exit velocity, estimated in Venusian conditions between 130 and 210~m~s$^{-1}$ \citep{aireyExplosiveVolcanicActivity2015}. Typical volatile content is between 2 and 7~wt\% \citep{wallaceChapter7Volatiles2015}. The volatile content is fixed at 5~wt\% as a reference.
The composition of Venusian volcanic plumes is not constrained. \cite{gaillardTheoreticalFrameworkVolcanic2014} studied the importance of the degassing pressure on the composition of magmatic gases from water-rich basalts, and estimated that for Venus it would be depleted in water compared to Earth by more than one factor of magnitude, and composed predominantly of CO$_2$. 
In this study, the reference composition is set to water vapour, the majority gas in Earth's volcanic outgassing, but CO$_2$ as the main gas is tested. The hypsometry of Venus is monomodal and peaks for altitudes below 1~km \citep{smrekarVenusInteriorStructure2018}. As a reference, the altitude of the vent is set to 1~km.

\section{Comparison with previous studies}
\label{Comp}
To compare to previous studies \citep{glazeTransportExplosiveVolcanism1999,carazzoRiseTurbulentPlumes2008}, the heat capacity of the atmosphere is fixed at 835~J~kg$^{-1}$~K$^{-1}$ and the pyroclast heat capacity at 920~J~kg$^{-1}$~K$^{-1}$, a typical value for scoria \citep{strobergHeatTransferCoefficients2010}, a common type of basalt volcanic rock. No wind shear is imposed. 
In these studies, the exit temperature and exit velocity are fixed at respectively 1400~K and 270~m~s$^{-1}$. The radius, a proxy for the MFR, is a free parameter. In the FPLUME model, the MFR M$_0$, the exit temperature and exit velocity $v_0$ are imposed, and then the vent radius is estimated as $\sqrt{M_0 / (\pi \uprho_0 v_0)}$ with $\uprho_0$ the plume density at the vent. The comparison is therefore not straightforward, as a similar radius can mean different vent conditions.

\begin{figure}[!ht]
 \centering
   \includegraphics[width=9cm]{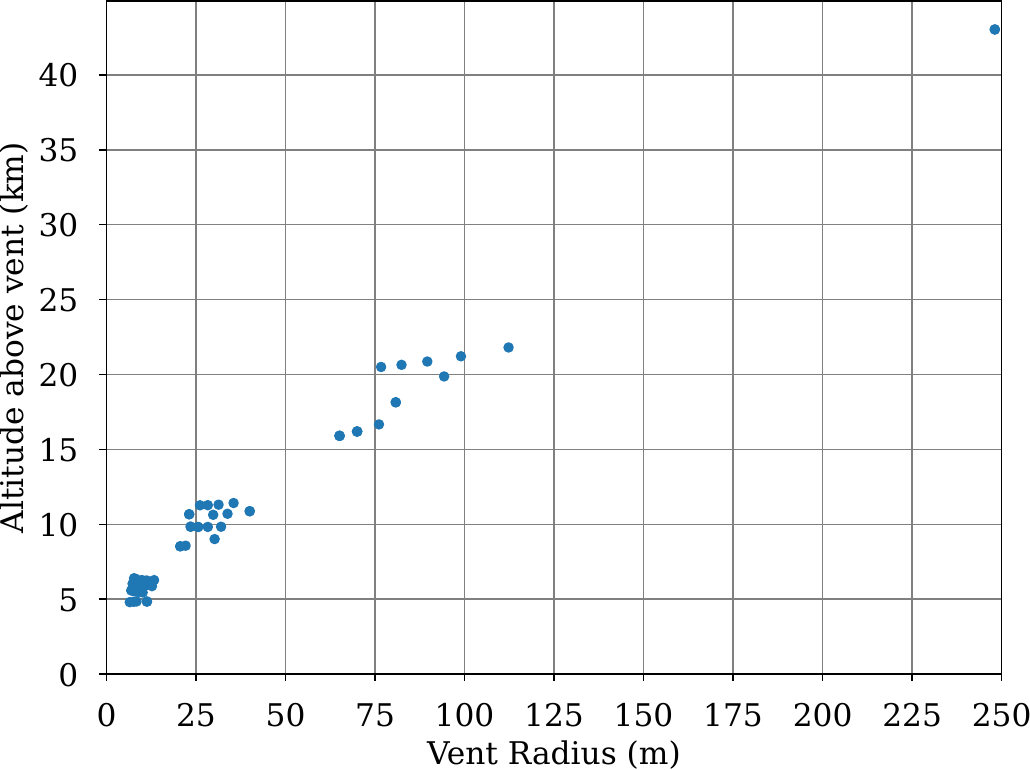}
    \caption{Maximum altitude reached by the plume with the FPLUME model in relation to the vent radius at the Equator for a constant atmospheric heat capacity of 835~J~kg$^{-1}$~K$^{-1}$ and a pyroclast heat capacity of 920~J~kg$^{-1}$~K$^{-1}$ and without wind shear.}
  \label{31}
\end{figure}

Figure~\ref{31} shows the relation between the maximum altitude reached by the plume and the radius at the vent obtained with the FPLUME model. Most selected plume scenarios have a vent below 125~m and an altitude below 25~km. One plume reaches above 40~km for a vent of 250~m. The relation between the radius of the vent and the plume maximum altitude is consistent with past studies \citep{glazeTransportExplosiveVolcanism1999,carazzoRiseTurbulentPlumes2008}.
However, there is a degeneracy of the relation between vent radius and plume height, with some plumes reaching the same altitude but with conditions at the vent that are different

\begin{figure}[!ht]
 \centering
   \includegraphics[width=\textwidth]{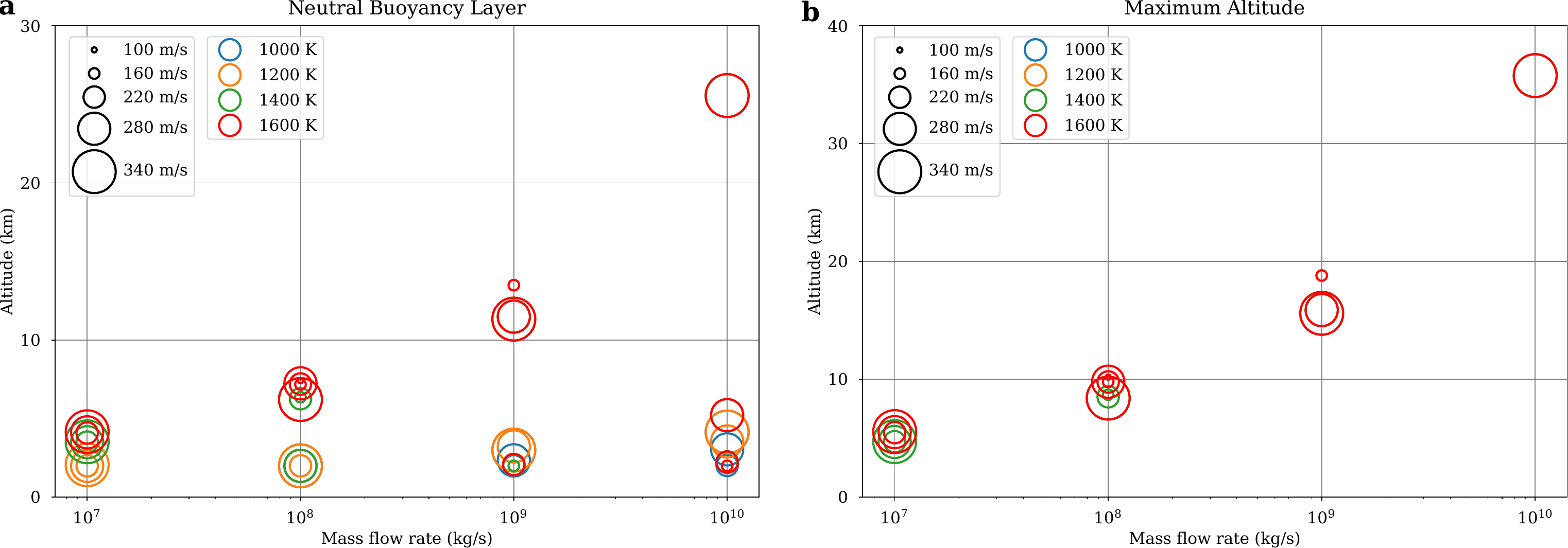}
    \caption{Plume neutral buoyancy altitude (left) and maximum altitude (b) in relation to the MFR, exit temperature and exit velocity at the Equator for a constant atmospheric heat capacity of 835~J~kg$^{-1}$~K$^{-1}$ and a pyroclast heat capacity of 920~J~kg$^{-1}$~K$^{-1}$ and without wind shear.}
  \label{32}
\end{figure}

Figure~\ref{32} shows plume neutral buoyancy altitude (a) and maximum altitude (b) in relation to the MFR, exit temperature and exit velocity for this case. There are three cases of plume: (1) total collapsing plume that do not reach the NBL, (2) Partial collapsing plumes that reach the NBL and then collapse, (3) Stable plumes that reach the NBL with an excess of momentum and the plume elevates above into the umbrella region \citep{costaDensitydrivenTransportUmbrella2013}. The plumes in case (1) have no points in Figure~\ref{32}. Plumes in case (2) only have points in Figure~\ref{32}-left, and plumes in case (3) have points in both Figure~\ref{32} panels. The sum of the cases (2) and (3) forms a group where plumes are reaching NBL and are named (pseudo)stable plumes hereinafter. Around 62\% of the cases have a neutral buoyancy altitude, but only 26\% have a maximum altitude. These late cases are stable plumes, whereas the others are collapse plumes. These collapse points are mainly for MFR superior to 10$^{9}$~kg~s$^{-1}$, very large plumes necessitating subsequent energy to rise. From the three vent parameters, the MFR has the biggest impact on the plume maximum altitude. When it is increasing, the altitude reached by the plume increases. The exit temperature and exit velocity have a limited impact when the plume is either stable or partially stable, but increasing these two parameters increases the energy and allows additional plume to reach (partial)stability. Below 1400~K, no plume is stable. No plume reaches the cloud floor altitude (45~km), and only one plume reaches the 30-45~km VenSpec-H observational region. For plumes with an MFR below 10$^{7}$~kg~s$^{-1}$, 85\% are stable with a maximum altitude at 3.0~km. No plumes are stable for 10$^{10}$~kg~s$^{-1}$, but two-thirds are collapse plumes reaching 6~km, with the possibility of co-ignimbrite plumes.

\section{A realistic Venusian environment}
\label{Real}
\subsection{Variable atmospheric heat capacity}

The specific heat of the environment plays a key role in the buoyancy of a volcanic plume. On Earth, it is considered constant, varying with temperature at 5\% at most. Whereas on Venus, the atmospheric specific heat varies by a third from 738~J~kg$^{-1}$~K$^{-1}$ at 100~km of altitude to 1181~J~kg$^{-1}$~K$^{-1}$ near the surface \citep{seiffModelsStructureAtmosphere1985}. Heat capacity C$_p$ can be approximated by the expression \citep{lebonnoisSuperrotationVenusAtmosphere2010} 
\begin{equation}
    C_p (T) = C_{p0} \cdot (\frac{T}{T_0})^\upnu 
\end{equation}
with $C_{p0}$ = 1000~J~kg$^{-1}$~K$^{-1}$, $T_0$ = 460~K and $\upnu$
 = 0.35
 \bigbreak
To account for the temperature dependency of the atmospheric heat capacity, the calculation of the specific enthalpy of the atmosphere at the temperature T$_a$ in the model is now calculated as follows:
\begin{equation}
\label{eq1}
H(T_a) =  \int_{0}^{T_a} C_{p}(T) \,dT = 86.6365 \cdot T_a^{1.35} 
\end{equation}

\begin{figure}[!ht]
 \centering
   \includegraphics[width=11cm]{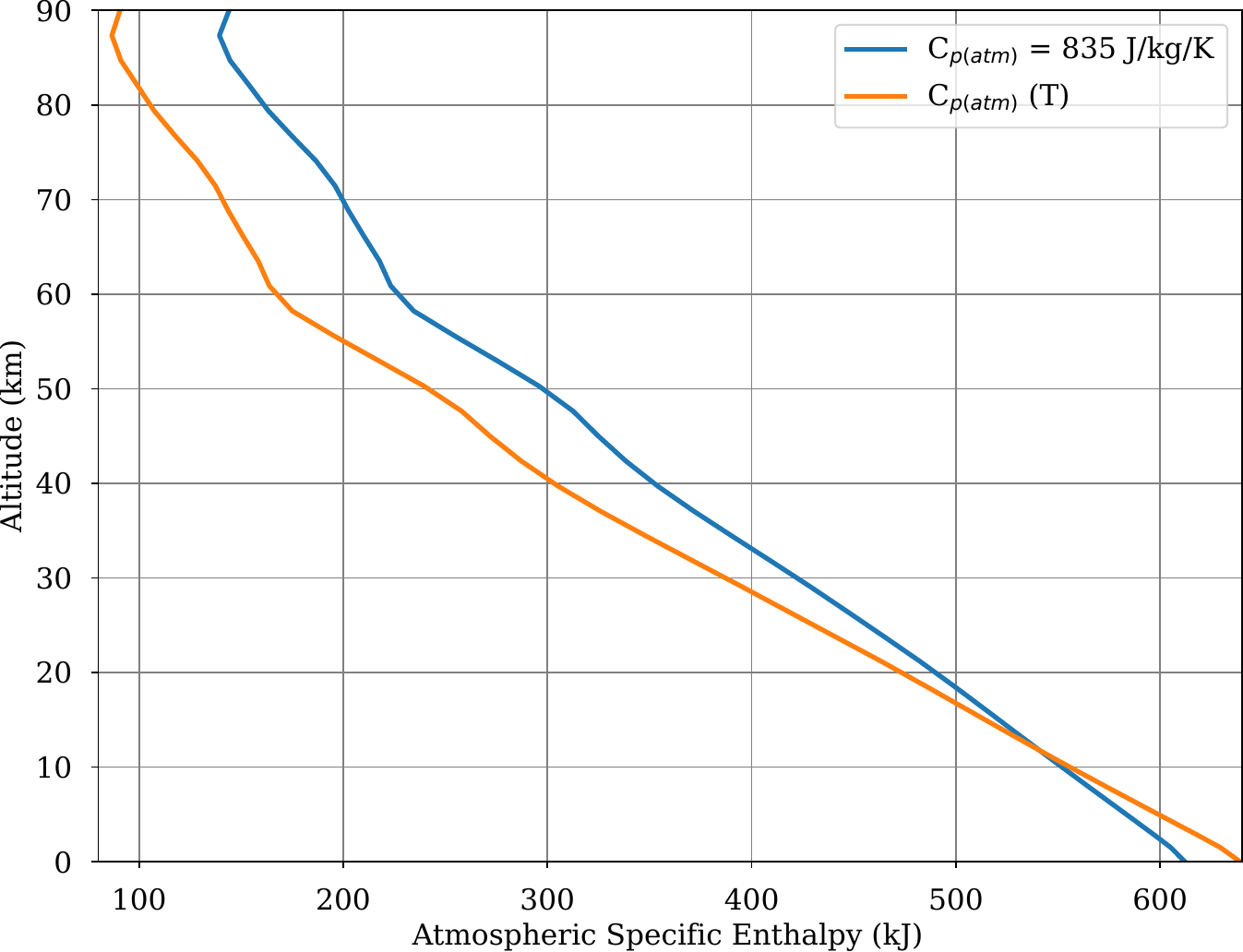}
    \caption{Vertical profiles of the Venusian atmospheric specific enthalpy (KJ) for a constant heat capacity (blue) and following the expression in equation~\ref{eq1} (orange).}
  \label{41}
\end{figure}

\bigbreak
Figure~\ref{41} shows the vertical profile of the atmospheric specific enthalpy for the two atmospheric heat capacity cases. The values close to the surface are stronger by the constant value by 4\%, but lower by 20\% at 50~km and by 30\% at 70~km.

\begin{figure}[!ht]
 \centering
   \includegraphics[width=\textwidth]{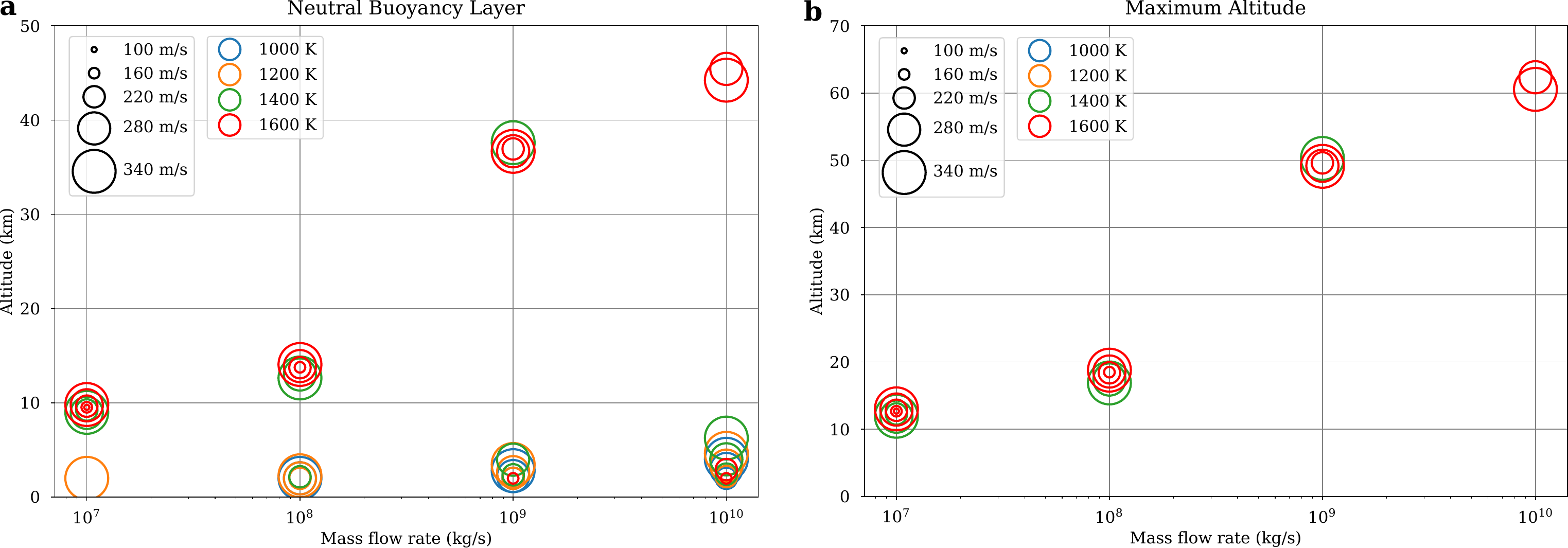}
    \caption{Plume neutral buoyancy altitude (a) and maximum altitude (b) in relation to the MFR, exit temperature and exit velocity at the Equator for an atmospheric heat capacity varying with temperature and a pyroclast heat capacity of 920~J~kg$^{-1}$~K$^{-1}$ and without wind shear.}
  \label{412}
\end{figure}

Figure~\ref{412} shows the plume neutral buoyancy altitude (a) and maximum altitude (b) for the temperature varying atmospheric heat capacity. Compared to Figure~\ref{32}, there is a significant increase in the altitude reached by the plume. Due to a lower atmospheric specific enthalpy above 10~km, 7\% of the volcanic plumes reach cloud altitudes. However, there is a slight decrease in the number of (pseudo)stable plumes and stable plumes, respectively with 58\% and 25\% of the cases. This is due to the increase in atmospheric specific enthalpy below 10~km. This new profile of atmospheric specific enthalpy will favour energetic plumes, increasing their maximum altitude.

\subsection{Wind shear}

\begin{figure}[!ht]
 \centering
   \includegraphics[width=\textwidth]{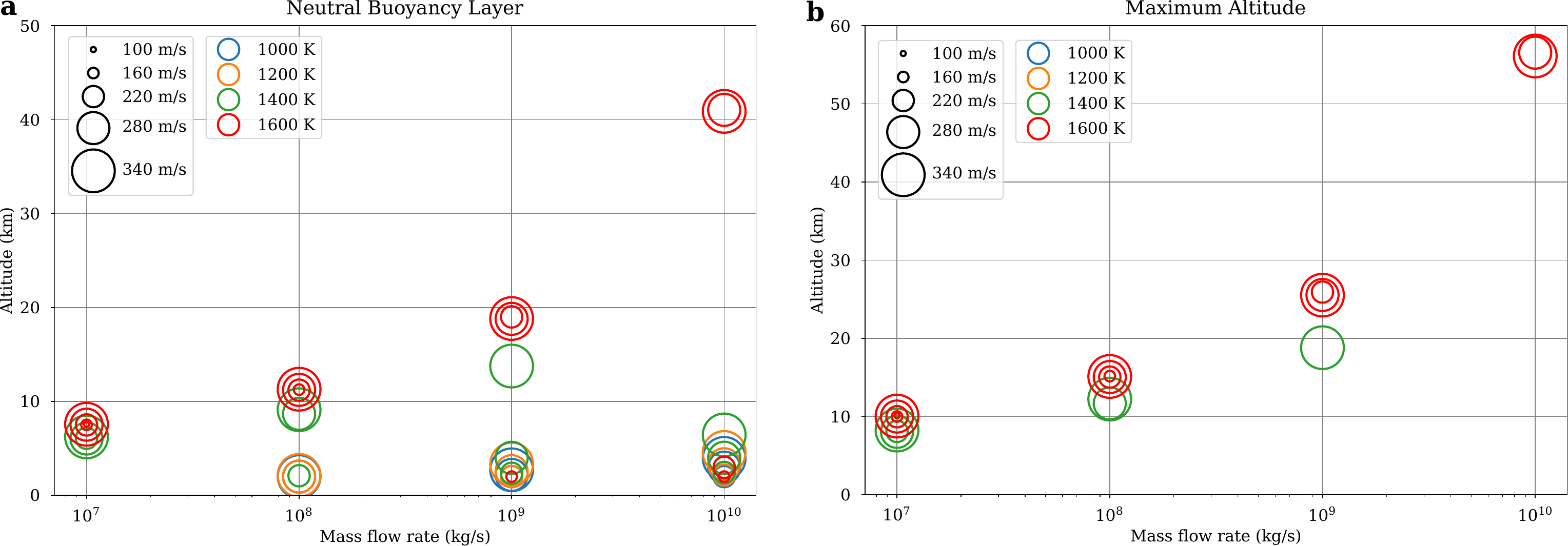}
    \caption{Plume neutral buoyancy altitude (a) and maximum altitude (b) in relation to the MFR, exit temperature and exit velocity at the Equator with wind shear and for an atmospheric heat capacity varying with temperature and a pyroclast heat capacity of 920~J~kg$^{-1}$~K$^{-1}$.}
  \label{42}
\end{figure}

On Earth, wind shear causes enhanced entrainment of air and horizontal momentum, plume bending, and a decrease in plume rise height \citep{bursikEffectWindRise2001}. On Venus, the atmosphere is in a super-rotation state, where in the clouds, the atmosphere circles the planet 60 times faster than the solid body. The zonal wind strictly increases from the surface to 80~km \citep{sanchez-lavegaAtmosphericDynamicsVenus2017}, reaching values up to 120~m~s$^{-1}$ at cloud-top altitudes in the Equator and around 90~m~s$^{-1}$ at high latitudes. Meridional wind is only a few metres per second below 80~km and can therefore be neglected. The impact of such shear on the plume dynamics has never been estimated for Venus.
\bigbreak
Figure~\ref{42} shows the plume neutral buoyancy altitude (a) and maximum altitude (b) with wind shear profile displayed in Figure~\ref{21}-left and a temperature varying atmospheric heat capacity. The number of stable plumes is the same, 25\%, but the number of (pseudo)stable plumes is slightly lower at 56\%. Moreover, there is a strong decrease in the altitude reached by the plumes. 87\% of the stable plumes do not exceed 15~km against 85\% without wind shear, and only 2.5\% of the cases reach above the 45~km cloud altitude range, against 7\% without wind shear. For plumes with an MFR below 10$^{7}$~kg~s$^{-1}$, 70\% are stable with a maximum altitude at 6~km. No plumes are stable for MFR superior to 10$^{10}$~kg~s$^{-1}$.

This wind shear bending angle will depend on the plume strength \citep{aubryImpactsClimateChange2019}. The stable plume angle comprises between 35 and 71$^{\circ}$ for the stable plumes, 90$^{\circ}$ corresponding to a plume perpendicular to the ground. The highest value is obtained for plumes reaching the clouds, i.e. the most energetic plumes.
\bigbreak
The wind will also affect the plume radius. With wind, the maximum plume radius, reached at the neutral buoyancy region, ranges from roughly 3 to 26~km, the majority below 10~km. Whereas without the presence of wind shear, no plume exceeds a 20~km radius.

\section{Vertical propagation}
\label{Vert}

Explosive volcanic plume propagation can be decomposed into three distinct regions: a jet phase, a convective phase and an umbrella phase \citep{woodsSustainedExplosiveActivity2013}. At the vent exit, the momentum-driven jet typically extends to 1-2 km on Earth, where the gas-particle mixture undergoes rapid decompression to atmospheric pressure, producing significant expansion as a highly turbulent flow. As the flow ascends, it entrains atmospheric gases, thus reducing the average density of the mixture. This is the convective region, where positive buoyancy drives the mixture up to the so-called neutral buoyancy level. At this altitude, the mixture becomes denser than the surrounding air, overshooting the level of neutral buoyancy, and decelerates until reaching a maximum height and spreads laterally, this is the umbrella region.
\begin{figure}[!ht]
 \centering
   \includegraphics[width=11cm]{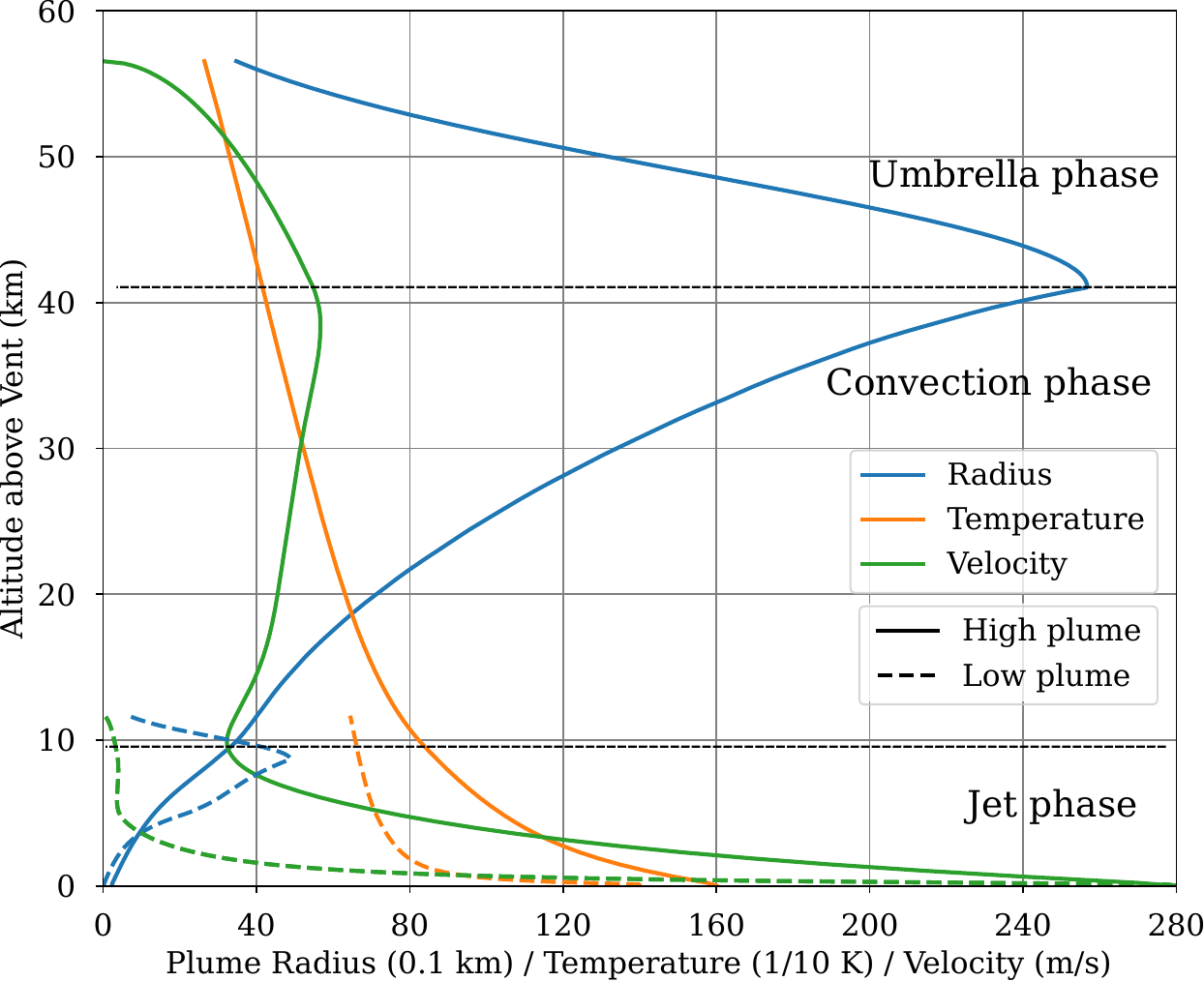}
    \caption{Vertical profiles of the plume radius (blue), temperature (orange), and velocity (green) for two stable cases: one high plume (solid lines) with an MFR of 10$^{10}$~kg~s$^{-1}$, an exit velocity 280~m~s$^{-1}$ of and an exit temperature of 1600~K, and a low plume (dotted lines) with an MFR of 10$^{8}$~kg~s$^{-1}$, an exit velocity 280~m~s$^{-1}$ of and an exit temperature of 1400~K. All cases are at the Equator with wind shear and for an atmospheric heat capacity varying with temperature and a pyroclast heat capacity of 920~J~kg$^{-1}$~K$^{-1}$.}
  \label{5}
\end{figure}
Figure~\ref{5} shows the vertical profiles of the plume radius, temperature and velocity for two stable cases, one 'high' plume reaching the clouds and one 'low' plume limited to the deep troposphere. The three phases are visible in the velocity. The jet phase corresponds to the high decrease from the surface to 5~km for the low plume and 8~km for the high plume, thicker than its Earth counterpart. Above, when the vertical gradient of the velocity is low, there is the convective region. And above, when the velocity is decreasing to zero, there is the umbrella region. 
The umbrella radius reaches a radius of 24~km and 4~km for the high and low plume cases, modest compared to Earth plume with similar height \citep{constantinescuRadiusUmbrellaCloud2021}.
In the model, the umbrella region is treated with a simple semi-empirical approximation with parameters based on Earth plumes \citep{folchFPLUME10IntegralVolcanic2016}. 3D simulations are therefore needed to test this approximation in Venusian conditions.
\bigbreak
During a volcanic event, some rock fragments, mineral crystals, and volcanic glass are also expelled with gas. As for other characteristics, on Venus, the composition and size of these pyroclasts are not known. On Earth, these materials can act as cloud condensation nuclei \citep{martucciImpactVolcanicAsh2012}, and can also lead to electrical discharge \citep{bennettMonitoringLightningApril2010}. On Venus, beside the thick cloud layer between 47 and 70~km, there is also the presence of a haze layer from roughly 35 to 47~km \citep{titovCloudsHazesVenus2018}, and a possible detached layer in the first 10~km \citep{griegerIndicationSurfaceCloud2004}. Charged aerosols were also measured below 35~km by VENERA-13 and 14 descent probes below \cite{lorenzDischargeCurrentMeasurements2018}. The role played by volcanic ash in those phenomena is not known. Figure~S1 presents the grain size distribution for the low and high plume cases using the second eruption event of El Chich\`on in 1983, a significant Plinian eruption \citep{sigurdsson1982EruptionsChichon1984}, as input.

\section{Sensitivity study}
\label{sens}
In this section, we explore the sensitivity of the plume stability to the characteristics of the plume itself.

\subsection{Plume volatile content}

The high Venusian atmospheric surface pressure would impact the volatile content needed to produce an explosive eruption \citep{headiiiVolcanicProcessesLandforms1986}, around few wt\% in current Venus surface conditions \citep{garvinMagmaVesiculationPyroclastic1982}. Estimations were conducted for the Scathach Fluctus pyroclastic flow feature \citep{ghailPyroclasticFlowDeposit2015}, and 4.5~wt\% if only H$_2$O, or 3\% H$_2$O plus 3\% CO$_2$ would be necessary to produce an explosive eruption \citep{aireyExplosiveVolcanicActivity2015}. To test the impact of the volatile content on the plume propagation, the volatile content was set to 2 and 7~wt\% for only water vapour, typical values at both ends of the range \citep{wallaceChapter7Volatiles2015}. The volatile content is inversely proportional to the exit temperature \citep{lesherChapter5Thermodynamic2015}. Therefore, for the 2~wt\% case, only 1400 and 1600~K, and for the 7~wt\% case, only 1000 and 1200~K represent a realistic temperature range.

\begin{figure}[!ht]
 \centering
   \includegraphics[width=\textwidth]{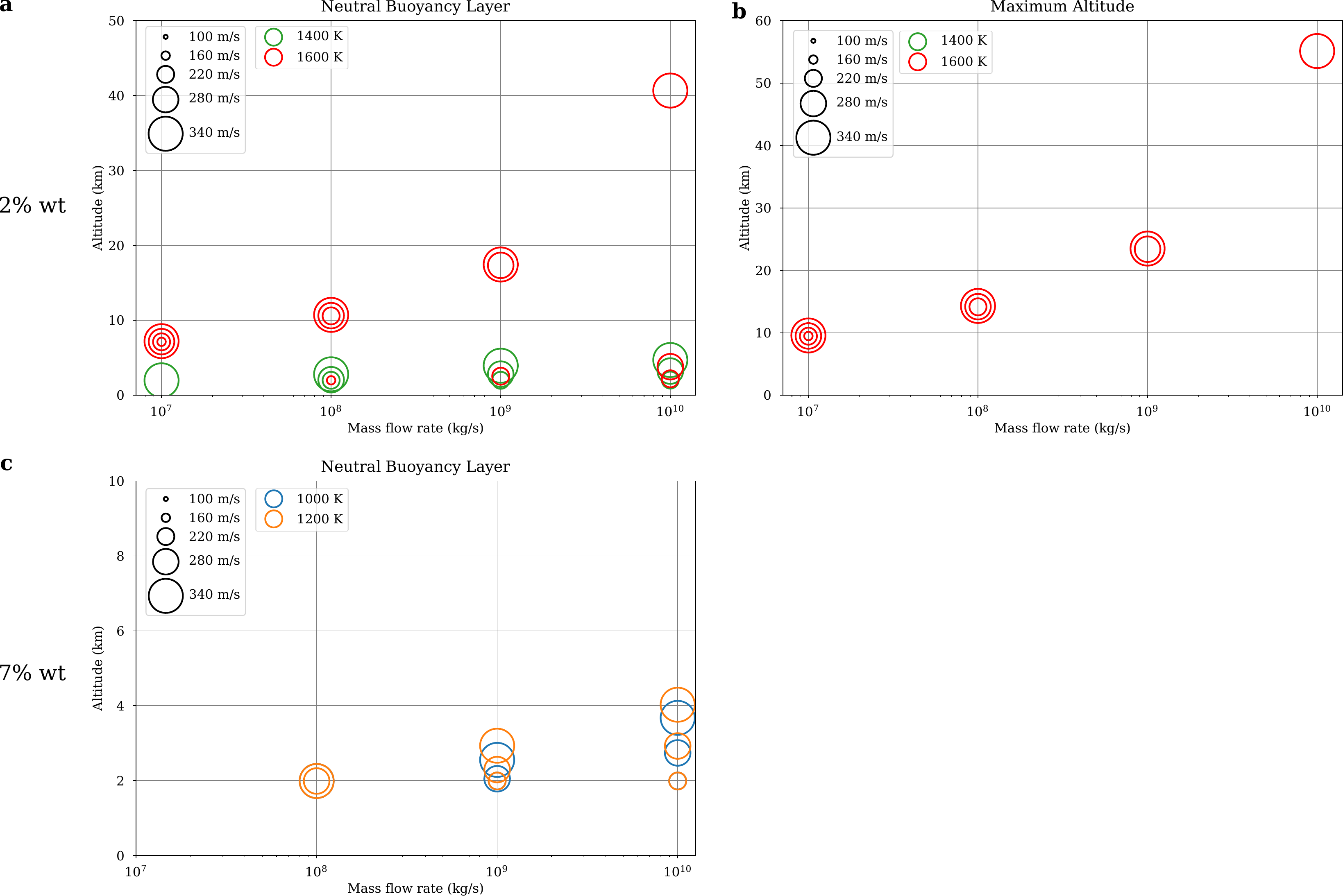}
    \caption{Plume neutral buoyancy altitude (left column) and maximum altitude (right column) in relation to the MFR, exit temperature and exit velocity for a volatile content of 2 (top row) and 7~wt\% (bottom row) at the Equator with wind shear and for an atmospheric heat capacity varying with temperature and a pyroclast heat capacity of 920~J~kg$^{-1}$~K$^{-1}$.}
  \label{43}
\end{figure}

Figure~\ref{43} shows the plume neutral buoyancy altitude (left column) and maximum altitude (right column) for a volatile content of 2 (top row) and 7~wt\% (bottom row), with a wind shear profile displayed and a temperature varying atmospheric heat capacity. With a decrease of volatile content, there is a slight decrease in the number of (pseudo)stable plumes compared to Figure~\ref{42} for the same exit temperature range, from 72 to 60\%, and the number of stable plumes drops by half. The maximum altitude range is of the same order of magnitude, with the same 2\% of plumes reaching the clouds, but with a smaller sample. For the 7~wt\% case, with the conservative low exit temperature range, inferior 1400~K, no plume is stable, but there is a slight increase in the number (pseudo)stable plumes, from 40 to 45\%. 33\% of plumes are stable by extending the temperature range to 1600~K.

\subsection{Plume volatile composition}

To test the impact of a plume volatile composition different from Earth counterparts \citep{gaillardTheoreticalFrameworkVolcanic2014}, the gas composition is set to pure CO$_2$ by setting the molecular mass, heat capacity, triple point and the condensation/evaporation equations to carbon dioxide values, with a 5~wt\% volatile content. Potential condensation effects are neglected.

\begin{figure}[!ht]
 \centering
   \includegraphics[width=\textwidth]{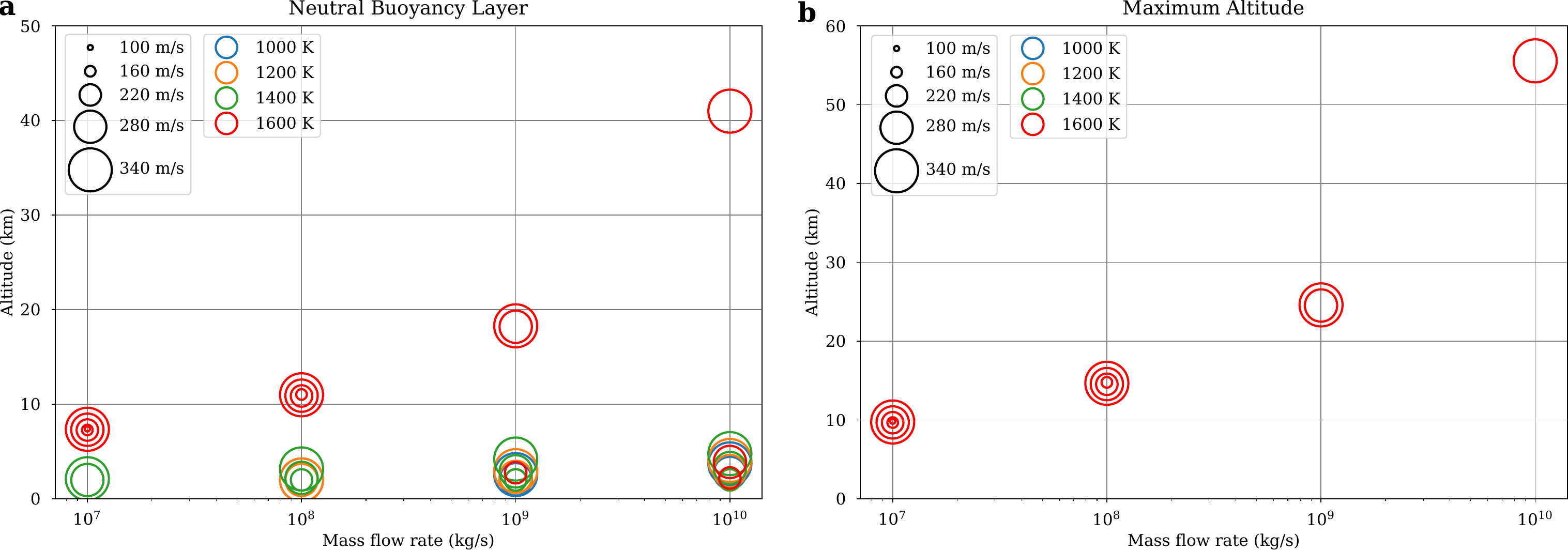}
    \caption{Plume neutral buoyancy altitude (a) and maximum altitude (b) in relation to the MFR, exit temperature and exit velocity for volatiles composed of CO$_2$ with a 5~wt\% volatile content at the Equator with wind shear and for an atmospheric heat capacity varying with temperature and a pyroclast heat capacity of 920~J~kg$^{-1}$~K$^{-1}$.}
  \label{432}
\end{figure}

Figure~\ref{432} shows the plume neutral buoyancy altitude (a) and maximum altitude (b) for volatiles composed of CO$_2$ with a 5~wt\% volatile content. Compared with a water plume (Figure~\ref{42}), the ratio of (pseudo)stable plumes decreases to 51\% while the ratio of stable plumes drops by almost a factor of two, 15\% of the cases. The number of plumes reaching the clouds is also decreasing by a factor of two. This is due to the lowest specific heat of CO$_2$ vapour, more than half the value for water vapour, resulting in less energetic plumes. 

\subsection{Pyroclast heat capacity}

The buoyancy of a plume is strongly controlled by its thermal energy and thus the characteristics of the pyroclast. As already mentioned, the composition of the plume on Venus is not known, and therefore the nature of the pyroclast is also unknown. In previous sections, the heat capacity of the pyroclast was set to 920~J~kg$^{-1}$~K$^{-1}$, on the lower edge of Earth values. To quantify the impact of the pyroclast heat capacity, it was set to 1600~J~kg$^{-1}$~K$^{-1}$, a typical high value used in such modelling on Earth \citep{woodsFluidDynamicsThermodynamics1988}. 

\begin{figure}[!ht]
 \centering
   \includegraphics[width=\textwidth]{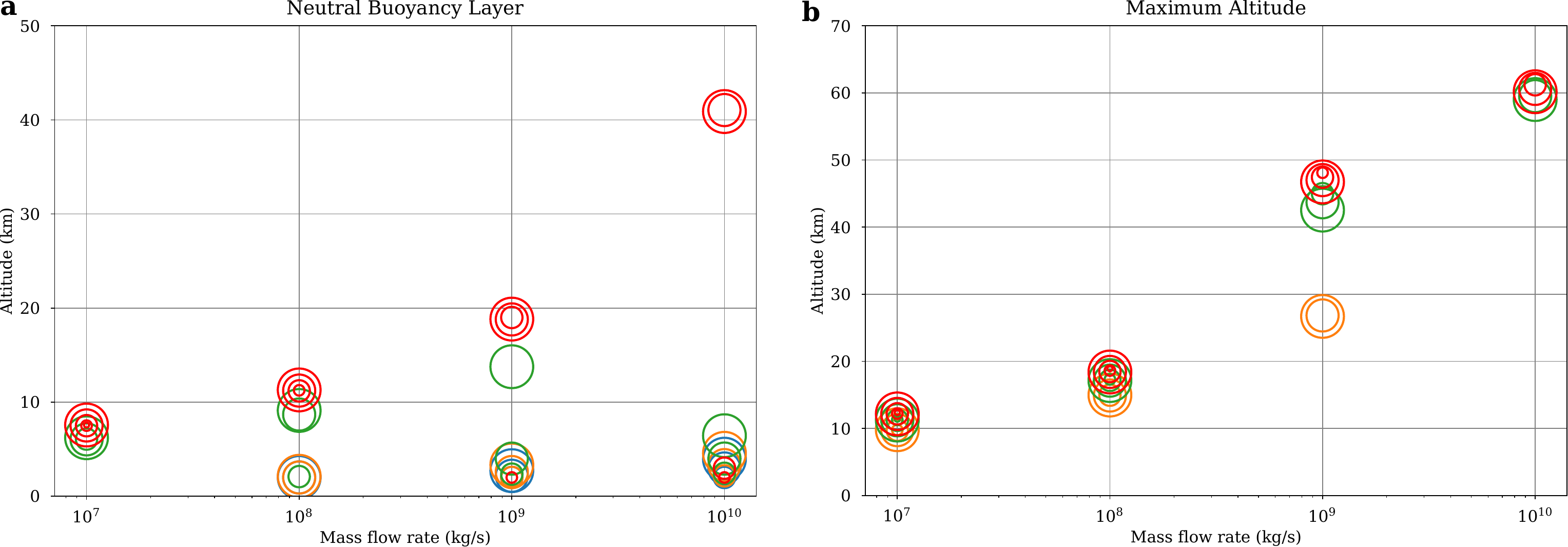}
    \caption{Plume neutral buoyancy altitude (a) and maximum altitude (b) in relation to the MFR, exit temperature and exit velocity with a pyroclast heat capacity of 1600~J~kg$^{-1}$~K$^{-1}$ at the Equator with wind shear and for an atmospheric heat capacity varying with temperature.}
  \label{44}
\end{figure}

Figure~\ref{44} shows the plume neutral buoyancy altitude (a) and maximum altitude (b) with the high pyroclast heat capacity, a wind shear profile displayed and a temperature varying atmospheric heat capacity. As expected, the number of stable plumes doubles compared to Figure~\ref{42}, and for the first time, plumes are stable with an exit temperature as low as 1200~K, due to the increased plume energy. 68\% of the cases are (pseudo)stable plumes with this high pyroclast heat capacity value against 56\% with the reference 920~J~kg$^{-1}$~K$^{-1}$ value, with twice the stable plume number, around 50\%. The maximum height reached by the plume is also impacted. The number of plumes reaching above cloud-floor altitudes is multiplied by four, above 11\% of the total cases. With a realistic Venus environment, plumes with an MFR of 10$^9$~kg~s$^{-1}$ can reach the cloud base. 56\% of the cases for plume with an MFR below 10$^{7}$~kg~s$^{-1}$ are stable, with a maximum amplitude of 8~km. With this nature of pyroclast, 2\% of plumes would reach the 30~km VenSpec-H limit against 10\% previously.

\subsection{Latitude of emission}

Volcanoes are present at every latitude on the surface of Venus \citep{hahnMorphologicalSpatialAnalysis2023}. In Figure~\ref{21}, the temperature at high latitudes exhibits a similar temperature profile below the clouds, and then decreases faster up to 62~km, and slightly increases above. This high-latitude thermal profile is colder and closer to the adiabatic lapse rate below 62~km than at the Equator. The zonal wind is weaker close to the pole, reaching 90~m~s$^{-1}$ at 64~km and decreasing above. These differences compared to the Equator are due to the presence of the warm polar mesosphere on top of the cold collar, observed by Pioneer Venus \citep{taylorStructureMeteorologyMiddle1980}, enclosing a warm polar vortex in each pole of Venus. 
\begin{figure}[!ht]
 \centering
   \includegraphics[width=\textwidth]{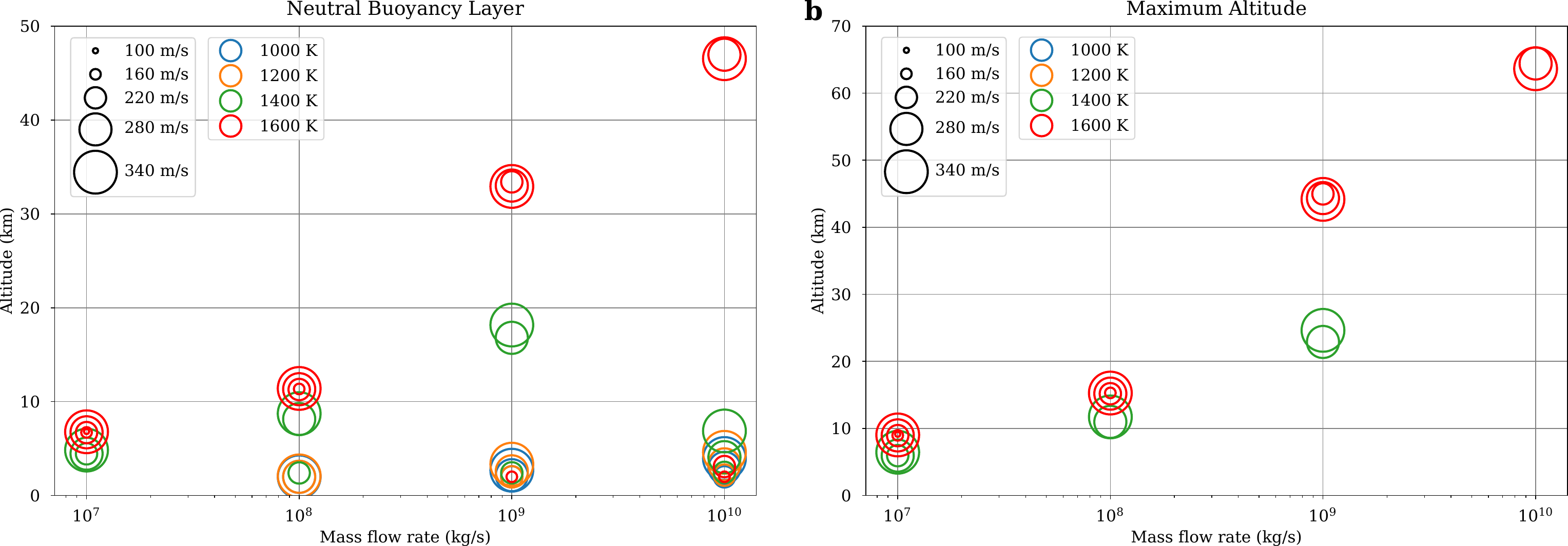}
    \caption{Plume neutral buoyancy altitude (a) and maximum altitude (b) in relation to the MFR, exit temperature and exit velocity at 75$^{\circ}$ of latitude with wind shear and for an atmospheric heat capacity varying with temperature and a pyroclast heat capacity of 920~J~kg$^{-1}$~K$^{-1}$.}
  \label{45}
\end{figure}
Figure~\ref{45} shows the plume neutral buoyancy altitude (a) and maximum altitude (b) at 75$^{\circ}$ of latitude with a wind shear profile displayed and a temperature varying atmospheric heat capacity. Compared to the equivalent set-up at the Equator (Figure~\ref{42}), the number of stable plumes at high latitude is very similar to the equator, respectively 26\% against 25\%. The ratio of (pseudo)stable plumes is also similar. The number of plumes reaching cloud floor altitudes is similar, around 2\%, but the maximum altitude increases. The highest plumes extend to 64~km against 56~km at the Equator. At high latitudes, these plumes are even above the clouds, due to the cloud-top altitudes, around 63~km versus 71~km at the Equator \citep{titovCloudsHazesVenus2018}. Around 6\% of the cases, three times more than at the Equator, reach above the 0~km VenSpec-H multiple species observation altitude. This increase of maximum reached height with latitude is consistent with previous models \citep{glazeTransportExplosiveVolcanism1999}.

The range of the plume angle at high latitudes is slightly extended compared to the Equator, between 27 and 79$^{\circ}$, due to more plumes reaching higher altitudes and more stable plumes. Twice more plumes have an angle above 65$^{\circ}$. The high latitude plume radius is also larger than at the Equator, with twice as many radii superior to 20~km, and the maximum radius reaching 32~km.

\subsection{Altitude of emission}

One Venus, the highest topographical feature is at 10~km of altitude. Between the 0~m altitude reference and 10~km, the pressure drops from 92 to 50~bars and the density from 68 to 38~kg~m$^{-3}$ \citep{lebonnoisVenusClimateDatabase2024}. To test the impact of the altitude of emission, the vent altitude was set to 4 and 8~km.
\begin{figure}[!ht]
 \centering
   \includegraphics[width=\textwidth]{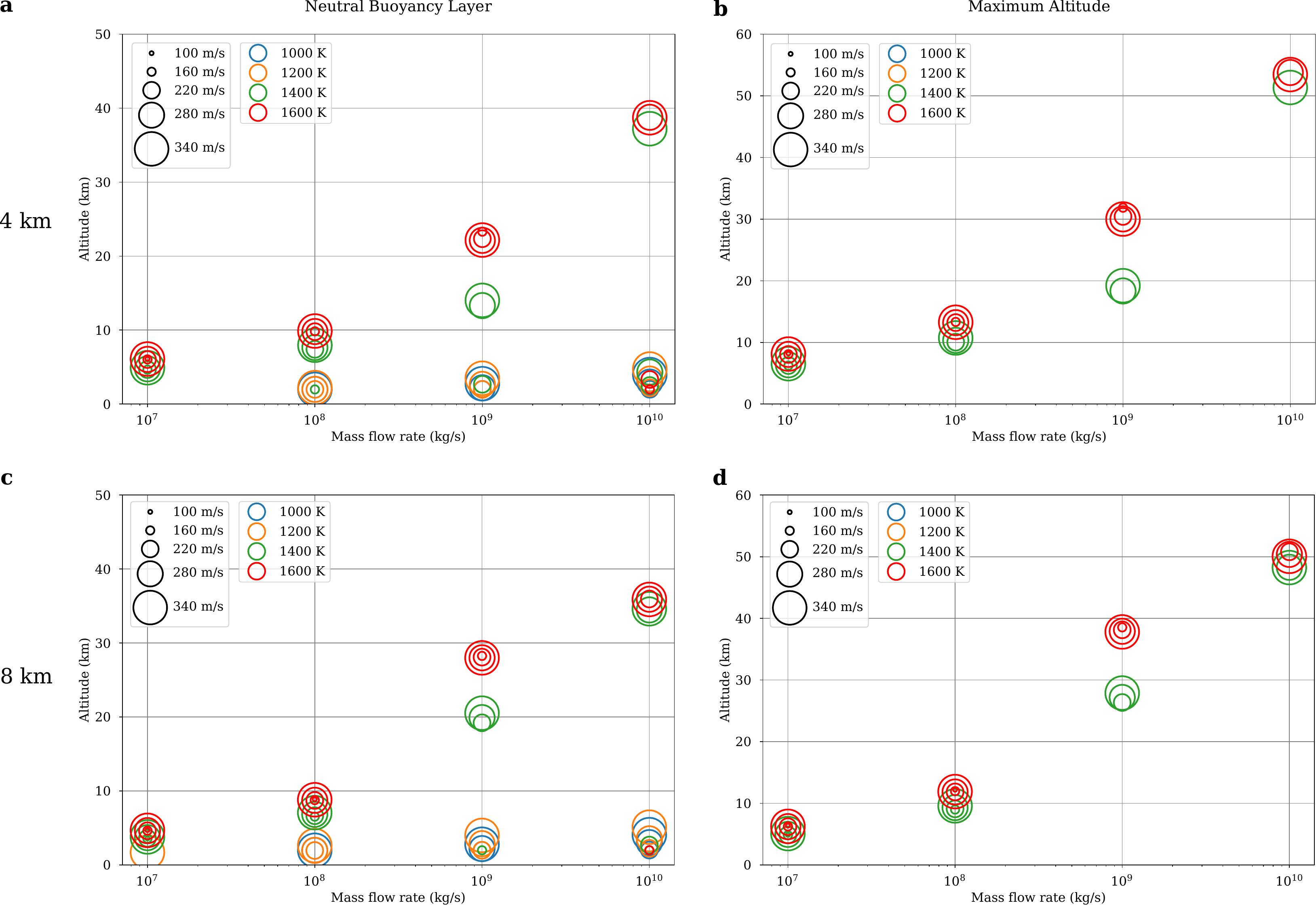}
    \caption{Plume neutral buoyancy altitude (left column) and maximum altitude (right column) in relation to the MFR, exit temperature and exit velocity at an elevation of 4~km (top row) and 8~km (bottom), both at the Equator with wind shear and for an atmospheric heat capacity varying with temperature and a pyroclast heat capacity of 920~J~kg$^{-1}$~K$^{-1}$.}
  \label{46}
\end{figure}
Figure~\ref{46} shows the plume neutral buoyancy altitude (left column) and maximum altitude (right column) for a vent at 4~km (top row) and 8~km (bottom row) with a wind shear profile and a temperature varying atmospheric heat capacity. With increasing the vent elevation, the pressure decreases and the number of (pseudo)stable plumes and stable plumes increases respectively from 56 to 66\% and from 25 to 38\% for a 1 and 8~km vent elevation. The plume ratio reaching the clouds also rises, from 2 to 6\%. For MFR below 10$^{7}$~kg~s$^{-1}$, the number of stable plumes steadily increases from 35 to 46\% for respectively a 1 and 8~km vent elevation. The altitude of the vent has only a limited impact on the maximum plume height. At higher altitudes of emission, 11\% of plumes would reach 30~km VenSpec-H multiple species observation range.

\section{Discussion}
\label{Disc}

The thermodynamics of the Venusian atmosphere is different from Earth, with a specific enthalpy that does not linearly increase with temperature. This discrepancy will tend to favour energetic plumes on Venus, increasing their maximum altitude. Contrary to Earth, where the height of the tropopause plays a major role in the plume's elevation and equatorial plumes are favoured \citep{aubryImpactsClimateChange2019}, the thermal and wind profiles favors higher plumes in the Venus environment. Volcanic vents at a high altitude would generate plumes in an environment with a density decreased by half at most, which would thus reach higher altitudes. High elevation volcanoes above mid-latitudes would be the target to observe explosive volcanism. VenSpec spectrometers will be monitoring tropospheric gas at those latitudes at night. As on Earth, the ubiquitous superotating venusian wind shear will tend to bend the volcanic and reduce the maximum height reached on Venus.

However, the actual volatile contents of Venus magmas are not known, and could be much lower that rates used in this study. The likelihood of explosive volcanism could then be compromised in the current Venus climate. The composition of the volatile itself is unknown and will affect the vertical propagation of the plume. At Venus' surface conditions, CO$_2$ could be the most outgassed volatile, and its thermodynamic properties negatively affect the height of the plume. The nature of the pyroclast and its thermodynamic properties also have a large impact on the altitude reached by the clouds. To fully assess the impact of contemporary explosive volcanism on Venus, there is a strong need for understanding the conditions and properties, and their variabilities, of the magma at the vent. 

In Table~\ref{T2} are summarized the impact of the plume vertical propagation of the parameters regarding the Venusian environment in Section~\ref{Real} or the plume itself in Section~\ref{sens}.

\begin{table}[!ht]
\center
\scriptsize
  \begin{tabular}{ll|c|c}
 %\hline
Related to & & \% of (pseudostable plumes, stable plumes) & (average NBL, maximum plume height) in km \\
 \hline 
 %\hline
\multirow{2}{*}{Atmosphere} & Atmospheric heat capacity & {\color{red}$\searrow$ (62,26) $\rightarrow$ (58,25)} & {\color{Green}$\nearrow$ (5,35) $\rightarrow$ (10,62)} \\
\cline{2-4} &  Wind shear & {\color{red}$\searrow$ (58,25) $\rightarrow$ (56,25)} & {\color{red}$\searrow$ (10,62) $\rightarrow$ (7,56)}  \\
 \hline
 \hline
\multirow{6}{*}{Plume} & Low volatile content & {\color{red}$\searrow$ (72,50) $\rightarrow$ (60,25)} & {\color{red}$\searrow$ (10,56) $\rightarrow$ (7,55) }\\\cline{2-4}
 & High volatile content & {\color{Green}$\nearrow$ (40,0) $\rightarrow$ (45,0)} & {\color{blue}$\sim$ (3,-) $\rightarrow$ (3,-) } \\\cline{2-4}
& Volatile composition & {\color{red}$\searrow$ (56,25) $\rightarrow$ (51,15)} & {\color{red}$\searrow$ (7,56) $\rightarrow$ (5,55)} \\\cline{2-4}
& Pyroclast heat capacity & {\color{Green}$\nearrow$ (56,25) $\rightarrow$ (68,50)}  & {\color{Green}$\nearrow$ (7,56) $\rightarrow$ (14,61)} \\\cline{2-4}
& Latitude of emission & {\color{Green}$\nearrow$ (56,25) $\rightarrow$ (56,26)}  & {\color{Green}$\nearrow$ (7,56) $\rightarrow$ (9,64)} \\\cline{2-4}
& Altitude of emission & {\color{Green}$\nearrow$(56,25) $\rightarrow$ (66,38)} & {\color{Green}$\nearrow$ (7,56) $\rightarrow$ (10,50)}  \\\cline{2-4}
 \end{tabular}
\smallbreak
\caption{Summary of the impact of the different parameters investigated in this study on the percentile of the (pseudostable plumes, stable plumes) in the middle column and on (average NBL, maximum plume height) in the right column. {\color{Green}{$\nearrow$}} means an increase, {\color{red}{$\searrow$}} means a decrease and {\color{blue}$\sim$} means marginal change}
\label{T2}
\end{table}

\section{Conclusion}
\label{Conc}

In this study, we used the eruption column model FPLUME to investigate the vertical propagation of a volcanic plume in the Venusian environment. In similar conditions, the model exhibits consistent plume height with previous studies. Previous studies have only investigated the plume propagation over limited range of vent parameters and with an idealised Venus atmosphere. 

Simulations were then performed with values for the mass flux rate, exit temperature, exit velocity and volatile content, covering a large range of Earth values and consistent with previous estimates for Venus. 

A realistic Venusian environment was also implemented. Contrary to the Earth, the atmospheric heat capacity strongly varies with temperature, and therefore with height. This effect is taken into account for the first time and greatly affects the plume height, where plumes will be tempered below 10 km and favoured above. For the first time, the effect of windshear is considered for Venus. The super-rotating wind has a significant effect on plumes by plume-bending. The number of stable plumes would increase and the maximum altitude would slightly decrease. In this realistic environment, 25\% of plumes are stable plumes and around 58\% can rise without total collapse. Only 2\% of the cases rise at the 45~km cloud-floor altitudes. The plumes are only stable for exit temperatures above 1400~K, a high value for the 5~wt\% volatile content. Around 2\% of the cases, for a very high MRF, would reach the 30~km VenSpec-H measurement altitude range and the 45~km cloud-floor altitudes. For more modest and likely conditions, with MFR $\leqslant$10$^ {6}$~kg~s$^{-1}$, plumes can reach a maximum altitude of 6~km.

The vertical structure of the plume can be estimated with the FPLUME model. The jet region of the plumes is thicker than on Earth, up to 8~km versus 2~km, whereas the umbrella radius is smaller than for Earth equivalent plumes. With the 1983 El Chich\`on explosive plume mass distribution, the model predicts that large volcanic plumes can transport 250~$\upmu$m particles into the cloud layer. The effect of such injections needs to be assessed.

The sensitivity of the stability of the plume to the characteristics of the plume itself was explored, and qualitative impacts of each parameter were assessed. The change of volatile content is narrowing the exit temperature range and therefore decreasing the number of stable plumes. For high volatile contents, the temperature range does not allow stable plumes. With CO$_2$ as a unique gas, there is a decrease in both the number of stable plumes and the maximum altitude, due to the thermodynamic properties of carbon dioxide leading to less energetic plumes. The pyroclast heat capacity, and therefore composition, has a strong impact on the plume height. With a higher value of this heat capacity, meaning a more energetic plume, there is a significant increase in the number of plumes reaching the 45~km cloud base. At high latitudes, plumes can reach higher altitudes due to a thermal profile closer to the adiabat in the clouds and a weaker windshear. There is also an increase in the number of stable plumes. This latitudinal variation is consistent with previous studies. There is a limited impact of the vent elevation on the height reached by the plume, but the decrease by almost half of the atmospheric density between 1 and 8~km of altitude increases the stability of plumes with lower energy. 
\bigbreak
3D modelling of volcanic plumes is needed to study high plume propagation, where discrepancies between 1D and 3D models are known due to the complex turbulent mixing \citep{costaResultsEruptiveColumn2016}. The possibility of co-ignimbrite plumes, the formation of large connecting plumes above pyroclastic density currents, generated by collapsing fountains, could also be assessed with this type of model. The impact of the shape of the vent, which might play a role on Venus \citep{glazeExplosiveVolcanicEruptions2011}, could also be studied. 
\bigbreak
Volcanic injection into the atmosphere potentially plays a role in climatic processes on Venus. While the majority of volcanism could be effusive or passive degassing, the present study shows that explosive volcanism would preferably reach altitudes up to 15~km above the vent. The majority of the plumes are reaching NBL, but only a fraction are stable. Plumes at high latitudes and from high mountains can propagate higher. Under certain conditions, large values for temperature, velocity and mass flux at the vent, plumes can reach the VenSpec-H multiple species observability region and into the clouds, but no plumes are reaching cloud-top altitudes at the Equator. The likelihood of these vent conditions needs to be assessed.

%%%%%%%%%%%%%%%%%%%%%%%%%%%%%%%%%%%%%%%%%%%%%%%
%
% DATA SECTION and ACKNOWLEDGMENTS
%
%%%%%%%%%%%%%%%%%%%%%%%%%%%%%%%%%%%%%%%%%%%%%%%

\section*{Open Research Section}
The simulation outputs used to obtain the results in this paper are available in the open online repository \citep{Lefe25data}.

\section*{Conflict of Interest}
The authors declare that they have no conflicts of interest relevant to this study.

\section*{acknowledgments}
The authors would like to thank the two anonymous reviewers who helped to improve this study. ML would like to thank Colin F. Wilson for insightful discussions. ML acknowledges receiving funding from the European Union's Horizon Europe research and innovation program under the Marie Sk\l odowska-Curie grant agreement 101110489/MuSICA-V.

\newpage
\appendix
\begin{appendices}
\section{Supporting Information}
The FPLUME model transports ash and takes into account aggregation. As grain size distribution, the one from the second eruption event of El Chich\`on in 1983, a significant Plinian eruption \citep{sigurdsson1982EruptionsChichon1984}, is used.

\begin{figure}[!ht]
 \centering
   \includegraphics[width=\textwidth]{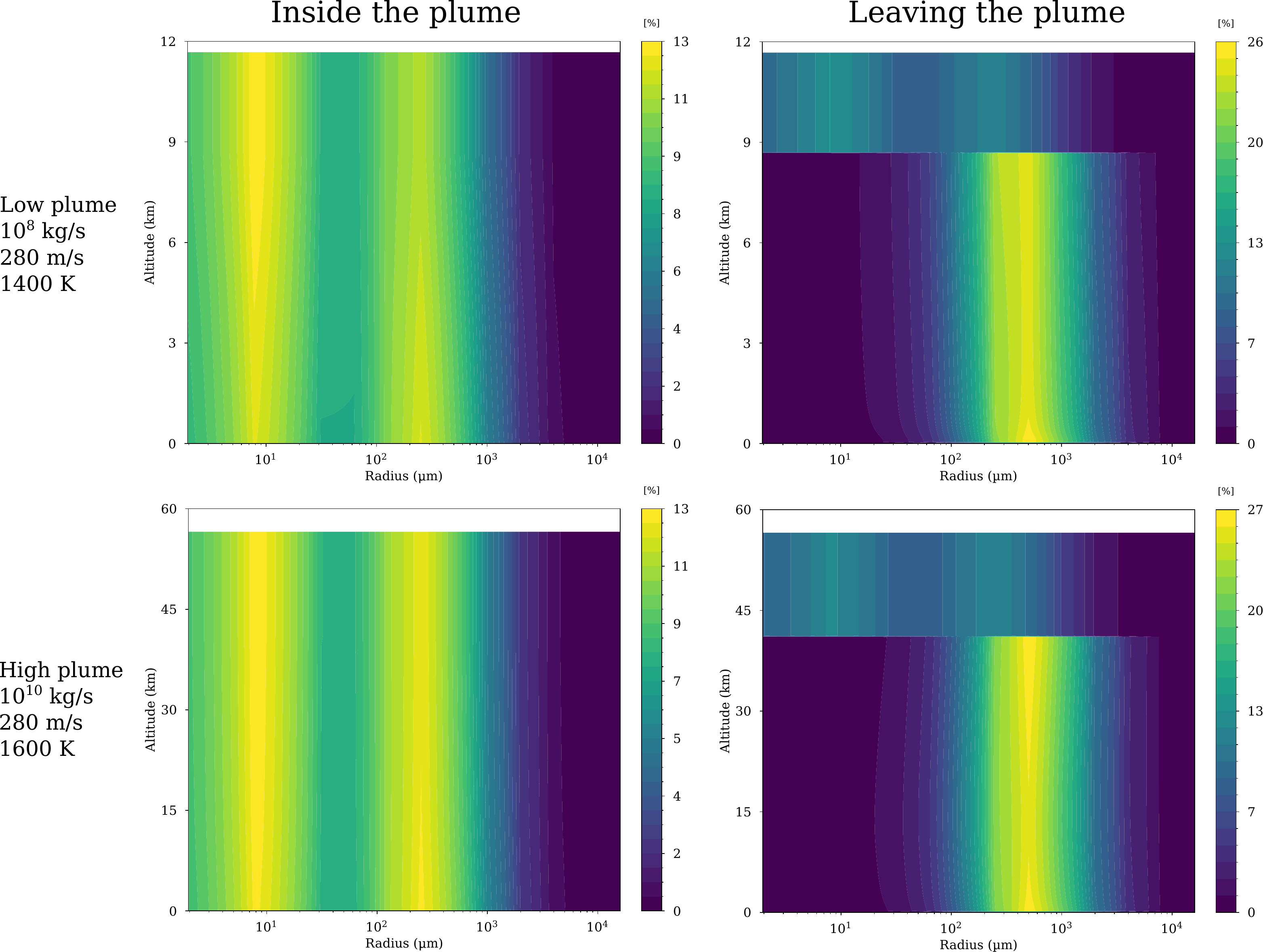}
    \caption{Mass distribution inside the plume (left) and leaving the plume (right) for the different particle sizes along the plume vertical extent for the low (top) and high (bottom) plume cases. Both cases are at the Equator with wind shear and for an atmospheric heat capacity varying with temperature and a pyroclast heat capacity of 920~J~kg$^{-1}$~K$^{-1}$.}
  \label{51}
\end{figure}

Fig~\ref{51} shows the vertical variation of the ash mass distribution inside the plume (left) and leaving the plume (right) for the low (bottom) and high (top) cases. Inside the plume, the bimodal distribution is visible throughout the plume vertical propagation with the strongest peak at 8~$\upmu$m and a second peak at 250~$\upmu$m. In the case of the low plume, this second peak decreases in intensity with altitude, whereas the first peak increases. The low plume loses the energy to transport the largest particle. In the case of the high plume, there is an equivalent decrease, but more tenuous.

The mass distribution of the material falling outside the plume is dominated by large particles of 500~$\upmu$m for both cases. In the umbrella region, the distribution changes drastically. It is more diffuse, with two small peaks at 8 and 250~$\upmu$m. The mass of the falling material represents between 10$^{-3}$ and 10$^{-4}$ of the mass inside the plume.
\bigbreak
With the example of the 1983 El Chich\`on explosive plume mass distribution, volcanic plumes are able to transport large particle ashes to the top of the troposphere and into the clouds.

\end{appendices}

\end{document}